\documentstyle[12pt,aps,prc,preprint,epsfig]{revtex}
\tightenlines

\begin{document}

\newpage
\title {
Strongly Enhanced Low Energy $\alpha$-Particle Decay in Heavy Actinide Nuclei
and Long-Lived Superdeformed and Hyperdeformed Isomeric States
}
\author{ A. Marinov$^{(1)}$, S. Gelberg$^{(1)}$, D. Kolb$^{(2)}$ and
J. L. Weil$^{(3)}$
}
\address{$^{(1)}$ Racah Institute of Physics, The Hebrew University, Jerusalem
91904, Israel}
\address{$^{(2)}$ Department of Physics, Kassel University, 34109 Kassel,
Germany}
\address{$^{(3)}$ Department of Physics and Astronomy, University of Kentucky,
 Lexington,
KY 40506, USA}
\maketitle
\begin{abstract}
Relatively low energy and very enhanced $\alpha$-particle groups have been
observed in various actinide fractions produced via secondary reactions in
a CERN
W target which had been irradiated with 24-GeV protons. In particular,
 5.14, 5.27 and 5.53
MeV $\alpha$-particle groups with corresponding half-lives of 3.8$\pm$1.0 y,
625$\pm$84 d and 26$\pm$7 d, have been seen in Bk, Es and Lr-No sources,
respectively. The measured energies are a few MeV lower than the known
ground state to ground state $\alpha$-decays in the corresponding neutron-
deficient actinide nuclei. The half-lives are
10$^{4}$ to 10$^{7}$ shorter than expected from the systematics
of $\alpha$-particle decay in this region of nuclei.
 The deduced evaporation residue cross sections are in the mb region, about
10$^{4}$ times higher than expected.
A consistent interpretation
of the
data is given in terms of production of long-lived isomeric states in the
second and third wells of the potential-energy surfaces of the parent
nuclei, which decay to the corresponding wells in the daughters.
The possibility that the isomeric states in the third minimum are actually
the true or very near the true ground states of the nuclei, and
consequences regarding the production of the long-lived superheavy
elements, are discussed.
\end{abstract}


\section{Introduction}

In a study of the radioactive decay of various actinide fractions \cite{1,2}
separated from
a CERN W target which had been irradiated with 24-GeV protons
 \cite{3,4,5}, very unusual phenomena which could not be understood
from nuclear systematics, had been
observed.

First \cite{2}, isomeric states with t$_{1/2}$ $\sim$ 0.6 y and  $\geq$  30 d
(10$^{4}$ - 10$^{5}$ times longer than the
 expected half-lives of the corresponding ground states) were found in
neutron-deficient
 $^{236}$Am and $^{236}$Bk nuclei, respectively.  About 3x10$^{5}$
atoms of $^{236}$Am and 4x10$^{4}$ atoms of $^{236}$Bk were produced
in the isomeric states, and decayed by the $\beta^{+}$ or electron capture
processes. The character of these states was not clear:
they are far from closed shells where high spin isomers are usually found, and
 they have very long lifetimes as compared to the known shape isomers.

Secondly, long-lived fission activities with half-lives of several years
\cite{1}, have
been seen in all the actinide sources from Am up to Fm, and perhaps also in
the Md-No and the No-Lr sources. Also here the origin of such long-lived
fission activities is not known.

Another observation was  very low energy, 3.0 and 4.0 MeV, particle groups,
which
 were seen in the
Am source in coincidence with L$_{\alpha1}$ X-rays in the Am region \cite{1}.
The  estimated lifetimes \cite{6} deviate by about 23 and 12 orders
of magnitude respectively from those obtained from the systematic
relationship between
$\alpha$-particle energies and their lifetimes \cite{7}. It therefore
was assumed \cite{6} that the observed particle groups were protons of
unknown origin rather than $\alpha$-particles.

Perhaps the most surprising phenomenon was the
 relatively low energy and very enhanced $\alpha$-particle
groups, observed in various actinide sources which were separated from
the W target \cite{1,2,6}.
For instance, a 5.14 MeV group with a half-life of about 4 y has been seen
in the Bk source, and a 5.53 MeV group with a half-life of about 26 d, has been
seen in the Lr-No source. One is faced here with the following
problems:

a) The long lifetimes of the nuclei as compared to corresponding ground state
systematics: It is expected and in fact
evident from the
identification of the $^{236}$Bk and $^{236}$Am nuclei, that neutron-deficient
actinide nuclei can be produced via the secondary reactions in the W target.
Lifetimes of several years for neutron-deficient nuclei around Bk, and of about
a month in the Lr-No region, are very much longer than the expected
 minutes or hours
in the
first case, and seconds in the second case. Here also one is led to the
conclusion that long-lived isomeric states of unknown character, which decay by
the observed $\alpha$-activities, were produced in the reaction.

b) The problem of the {\bf low} energy of the $\alpha$-particles: An energy of
5.14 MeV in the neutron-deficient Bk
region is a {\bf low} energy
as compared to typically 6-7 MeV, and 5.53 MeV is {\bf very low} for
Lr-No nuclei,
where the typical transition energies are above 8 MeV. The penetrability
factor for
6 MeV $\alpha$-particles in the Bk region is about 5 orders of magnitude
larger than for 5.14 MeV, and for 8 MeV in the No region is about 13 orders of
magnitude larger than for 5.53 MeV \cite{7}.  Naturally, one
assumes that an $\alpha$-decay from an isomeric state is of even higher
energy
than normal, because of the larger available transition energy. One is
therefore  faced with the problem,
 what is causing the nucleus
to decay with very low-energy $\alpha$-particles, when in principle
 much higher transition energy is available with correspondingly
 many orders of magnitude larger penetrability factor.

c) The problem of the
very enhanced
character \cite{6} of the $\alpha$-decay: According to
systematics \cite{7}, the half-lives for  5.14 MeV $\alpha$-particles for Bk
and nearby nuclei, and for 5.53 MeV
$\alpha$-particles in Lr-No, are around 10$^{5}$ to 10$^{6}$ y. The
 question
is how
the $\alpha$-particle decay rate can be {\bf enhanced} by 5 to 7 orders of
magnitude.

d) The problem of the large cross section of the heavy-ion reaction:
  The production yield in the
secondary reaction experiment depends on two cross sections: i) The cross
section for production of suitable fragments with high enough  kinetic energies,
which may interact with other W nuclei in the target, and lead to the formation
of the various actinide nuclei. ii) The heavy ion reaction, or the
 evaporation residue, cross section, which
depends first on the fusion cross section between such a
fragment and another W nucleus in the target, and secondly on the competition
with the fission process during the cooling of the produced compound nucleus.
 Using reasonable estimates about
the first process, evaporation residue cross sections in the region of mb
 are needed
in order to explain the production of about 10$^{4}$ actinide atoms
seen experimentally. These are unexpectedly large cross sections,
particularly when the production of nuclei in the Lr-No region is considered
where the typical evaporation residue cross sections are
below 1 $\mu$b \cite{8}.

Recently \cite{9,10,11} some similar abnormal decay phenomena have been
observed,
for the first time,
in several ordinary heavy-ion reaction studies.
In a study  of the $^{16}$O + $^{197}$Au reaction at
E$_{Lab}$ = 80 MeV \cite{9},
  an isomeric state has been found in
 $^{210}$Fr which decays by emitting a relatively low energy
  $\alpha$-particle group, with E$_{\alpha}$ = 5.20 MeV and
t$_{1/2}$ $\sim$ 90 m.   Since this half-life is longer than the known
half-life of the ground state of $^{210}$Fr (t$_{1/2}$ = 3.18 m),
it was concluded that a long-lived
 isomeric state had been formed in this nucleus. A t$_{1/2}$ of 90 m
for 5.20 MeV $\alpha$-particles in $^{210}$Fr is enhanced by a factor
 of 3x10$^{5}$ as compared to normal transitions \cite{7,9}. However, this
$\alpha$-group was observed in coincidence with $\gamma$-rays which fit
predictions for a superdeformed band \cite{9}.  Therefore the effect of large
deformations of the nucleus on the $\alpha$-particle decay was
calculated and found \cite{9} to be consistent with the observed enhancement.
It was argued \cite{9} that since the isomeric state decays to a high spin
state,
it should also have high spin, and since it decays by
 enhanced $\alpha$-particle emission
to state(s) in the second well of the potential, it should be in the second
 well itself.

In the same reaction  two long-lived proton activities
 with half-lives of about 6 h and 70 h were found \cite{10} with proton
energies of 1.5 -- 4.8 MeV with a  sharp line at 2.19 MeV. A possible
interpretation in terms of production of long-lived isomeric states in the
second well of the potential of the parent nucleus which decay by protons to
the normal states of the daughter was given \cite{10}.

Very recently \cite{11}, long-lived isomeric states,
 produced
by the $^{28}$Si + $^{181}$Ta reaction at bombarding energies of
125 and 135 MeV, which decay by strongly hindered
$\alpha$-particle and proton activities, have been discovered. In
particular, a very retarded 8.6 MeV $\alpha$-particle group with
40 d $\leq$ t$_{1/2}$ $\leq$ 2.1 y  (as compared to less than a 1
$\mu$s, typical for a normal transition of such high energy) was
found in coincidence with $\gamma$-rays of a superdeformed band.
This group has been
 consistently
interpreted \cite{11} in terms of production of a long-lived isomeric state
in the third
(hyperdeformed) well probably of $^{195}$Hg, which decays by very retarded
$\alpha$-particles, to the second (superdeformed) well in $^{191}$Pt.

In the present work it is shown that the low energy and very enhanced
$\alpha$-particle groups, seen in the heavy actinide, Bk, Es and Lr-No
sources,
can be consistently  interpreted as due to formation of long-lived isomeric
states in the second and third wells of the potential energy surfaces in the
parent nuclei, which
decay to the corresponding wells in the daughters. Furthermore, it is pointed
out that according to theoretical predictions \cite{12}, the third minima are
the actual ground states of the very heavy actinide nuclei.

\section{Experimental Procedure and Results}

The experimental procedure has been published before \cite{1,2,3,4} and will
be briefly reviewed.
Actinide fractions from Am up to Lr-No were separated \cite{1,2} from our (W3)
\cite{3,4} tungsten
target (6 cm long, 30 g, 99.95\% pure)
which was irradiated in CERN with about 1x10$^{18}$ ($\pm{20\%})$ protons of
24 GeV energy.\footnote{The chemical separation was performed by the late
A. M.~Friedman.} Fig.~1 shows a block diagram of the chemical procedure. The
main aim in the chemical separation was to separate the heavy actinides,
from Am up to Lr, from the rare-earth isotopes and from $\alpha$-emitting
nuclei such as Po, Ra, Ac, Th, U and Pu.
Several anion (Dowex-1) and cation (Dowex-50) exchange columns were used for
this purpose, and at the end two fractions were obtained. The first included
the remainder of the rare-earth elements, Am and part of the Cm. The second
fraction included the other part of Cm and all the heavier actinides. These
two fractions were then passed through Dowex-50 cation exchange columns with
ammonium $\alpha$-hydroxyisobutyrate acid at 87$^{o}$ C, in order to separate
individual actinides. The elution curves obtained from these last separations
are shown in fig.~2. The X-rays of the rare-earth elements, measured with
a Si(Li) detector were used throughout the whole procedure to identify the
contents of the fractions. Actinide sources were prepared as indicated in
fig. 2 using the measured elution positions of the rare-earth elements,
and known information on the actinides \cite{13}. The sources were prepared
on thin glass discs which were then heated to about 600$^{o}$ C to remove
any organic material present.

The decontamination factor for Th and Pu was estimated to be $\geq$ 10$^{8}$
\cite{2};
the decontamination factor from a typical rare earth element like Gd,
determined by a comparison of the residue number of $^{148}$Gd nuclei
seen in the Bk and Cf
sources \cite{1} with the total number of produced $^{148}$Gd nuclei,
 deduced from its known
production cross section \cite{14}, was about 9x10$^{4}$.
The intensity of $^{148}$Gd was found to be about 4000 times larger in the Cf
and Bk sources (where it should preferentially be) compared to the adjacent
Es and Cm sources \cite{1}.

$\alpha$-particle spectra were measured from the various sources
using 450 mm$^{2}$ Si surface barrier detectors. In the present paper
we concentrate on the 5.14, 5.27 and 5.53 MeV groups which have been seen with
the Bk, Es and Lr-No sources, respectively. The uncertainty in the energies
was about $\pm$40 keV. Fig.~3 (top) shows a typical $\alpha$-particle spectrum
obtained with the Bk source. (Fig. 1f of ref. [2] represents a corresponding
background spectrum). A pronounced peak at 5.14 MeV is seen in this
spectrum. (The other pronounced peak at 5.74 MeV was identified to be due
to the decay of
 $^{236}$Pu,
 obtained after several $\beta$$^{+}$ or EC transitions from an isomeric state
in $^{236}$Bk \cite{2}). Fig.~3 (bottom) gives the decay curve of the 5.14 MeV
group, measured during about 15 y. It is seen in this figure that the intensity
first grows with a half-life of 2.0$\pm$0.5 y and then decays with a half-life
of 3.8$\pm$1.0 y. Fig.~4 shows similar plots obtained with the
Es source, where a 5.27 MeV $\alpha$-particle group with a half-life of about
625 d is seen. In fig.~5 two spectra obtained with the Lr-No source are shown.
In the first spectrum (top) a pronounced peak at 5.53 MeV is seen, while in
the second one (bottom), which was measured about 3 months later, very little of
this peak
remained. From these two spectra a half-life of 26$\pm$7 d is obtained.
(Fig. 1e of ref. [2] is the corresponding background spectrum for fig. 4 (top)
and fig. 5).
From the measured intensities and lifetimes of the $\alpha$-particle groups,
it was estimated that the number of produced atoms were 3.3x10$^{4}$,
1.4x10$^{4}$ and 1.4x10$^{4}$, in the Bk, Es and Lr-No sources, respectively.
These number are correct to within $\pm$50\%.

\section{Discussion}

It is impossible to identify the above mentioned $\alpha$-particle groups
with any known activity in the whole nuclear chart. We shell now
show in some detail that one can rule out
the possibility that
the 5.14 MeV group is due to $^{208}$Po
(E$_{\alpha}$ = 5.12 MeV; t$_{1/2}$ = 2.898 y), the 5.27 MeV group is due
to $^{210}$Po (E$_{\alpha}$ = 5.30 MeV; t$_{1/2}$ = 138.4 d), and the
5.53 MeV group is due $^{222}$Rn (E$_{\alpha}$ = 5.49 MeV; t$_{1/2}$ = 3.83 d).

1) The 5.14 MeV group in fig. 3 is not due to $^{208}$Po since:

a) In the chemical procedure (fig. 1) there were 4 anion exchange columns
where Po
should have been strongly adsorbed.

b) Po is very volatile. The last stage in the preparation of the sources,
as mentioned above, was heating them to 600$^{o}$C. If some
Po was left after the chemical separation, then it would have evaporated
during this heating.

c) As mentioned above the half-life of $^{208}$Po is 2.9 years
and it does not have a long-lived
parent. If the 5.14 MeV is from $^{208}$Po, then the intensity at the beginning
in fig. 3 (bottom) should have been about twice as much as the intensity seen
at around 3 y, namely 160$\pm$12 and not 20$\pm$20 as seen. In
addition the intensity at the last point in fig. 3 (bottom) should have been
5$\pm$0.4 and not 20$\pm$6 as seen.

2) The 5.27 MeV group seen with the Es source in fig. 4 is not due to
$^{210}$Po since:

a)  (1a) and (1b) above.

b) The deduced half-life is 625$\pm$85 d which is about 4.5 times longer
than the known half-life
of $^{210}$Po of 138.4 d.

This group  is also not due to $^{210}$Po which is fed
from $^{210}$Pb because:

c) In the chemical separation procedure (fig. 1) there were one
 1.5M HCl anion exchange and three cation exchange columns where Pb should
have been adsorbed.

d) The half-life of $^{210}$Pb is 22.3 y and of $^{210}$Po is 138.4 d.
By normalizing to the data around 120 d and using mother-daughter relationship
for these half-lives one gets that the intensity
in the first
measurement shown in fig. 4 (bottom)  should have been 23$\pm$3
counts, and in the last measurement
290$\pm$32. These figures are in complete contradiction to the
observed data where the corresponding numbers are
 135$\pm$22  and 40$\pm$5, respectively.

3) The 5.53 MeV group seen with the Lr-No source
in fig. 5
is not due to $^{222}$Rn (which is introduced to the counters with air) for the
following reasons:

a) If the 5.53 MeV group is from $^{222}$Rn then groups from its
short-lived daughters
$^{218}$Po and $^{214}$Po, of about the same intensity, should have been seen
at 6.00 MeV  and 7.69 MeV  in fig. 5 (top). Nothing is seen in
these energy regions.

b) A group from the air should have been very narrow, about 25 keV, and not
about 130 keV, as seen experimentally in fig. 5.

c) One does not see such a group of similar intensity in other spectra which
were measured under identical conditions like the one displayed in fig. 5
(bottom), or in the relevant background spectrum which is shown in fig. 1e
of Ref. [2].

It is also not due to $^{222}$Rn which is fed from $^{226}$Ra, since
one does not see in fig. 5 (top) the 4.78 MeV $\alpha$-particle group
of $^{226}$Ra,
 the half-life of $^{226}$Ra is 1600 y  much
longer than the measured half-life of 26$\pm$7 d,
and also because of (3a)
above.\\

$\alpha$-particle decays with similar energies but with very different
half-lives of a few seconds are known in very neutron-deficient
Yb to Pt nuclei. For instance, $^{167}$Re and $^{180}$Pt decay with
$\alpha$-particles of about 5.14 MeV; $^{174m}$Ir, $^{174}$Ir,
$^{166}$Re, $^{170}$Os, $^{160m}$Ta, $^{158}$Hf, $^{154}$Yb and
$^{160}$Ta decay with $\alpha$-particles around 5.3 MeV, and $^{156}$Lu
with a 5.57 MeV group. In principle one may assume that the activities
seen by us are due to such decays which are fed by unknown long-lived
isomeric states. This possibility has been studied by us in some detail.
Six of the above mentioned isotopes, those above W,
 can in principle be produced by
either (p,xn) or secondary reactions, where the number of evaporated
neutrons is between 9 to 17. For instance, $^{167}$Re can be reached
by the $^{182}$W(p,16n)$^{167}$Re reaction, and $^{174}$Ir by
$^{182}$W($^{6}$Li,14n)$^{174}$Ir reaction. Additionally,
five of the above mentioned
isotopes, those below W,  can in principle be produced
by spallation reactions
where 12-13 neutrons have to be evaporated beyond the most neutron
deficient isotopes which are known to be produced by such reactions.
By using very modest assumptions about the production cross sections
of these nuclei,\footnote{For instance,
for the production of $^{167}$Re from the $^{182}$W(p,16n)$^{167}$Re reaction,
  a cross section of 1.5 mb was assumed
for the production of $^{178}$Re, like in the production of $^{177}$W from
$^{181}$Ta \cite{15}, and this was reduced by a factor of 5 for each additional
evaporated neutron (Fig. 3a in Ref. \cite{16}). This factor is very modest.
It may be much larger for very neutron-deficient nuclei where the proton
binding energy decreases and the neutron binding energy increases. (The
binding energy for a proton in $^{167}$Re is about 260 keV and for a
neutron it is about 11 MeV). For the secondary reactions, the cross sections
for the required high energy (above the corresponding Q-values)
  heavy ions, like $\alpha$-particles, $^{6}$Li and
$^{7}$Be, were estimated from Ref. \cite{17}, and evaporation residue cross
sections of 1 $\mu$b, 1 nb and 1 pb were assumed for 9n, 14n, and 16n
reactions, respectively. These are rather modest assumptions. For the
spallation reactions, known cross sections \cite{15,18} for the production of
the most neutron-deficient nucleus were used, and then reduced by a factor of 5
for each additional (12 or 13) evaporated neutron (see above).}
and very modest assumptions or
measurements of the decontamination factors of our sources from these
elements,\footnote{For Hf, Yb and Lu the decontamination factors were
 measured by a comparison
with the X-ray intensities from a 0.4\% of the solution of the LaF$_{3}$
precipitate before the main chemical separation took place (see Fig. 1).
For the other elements it was assumed that the decontamination factor due to
adsorption in an anion or a cation exchange column is at least
as large as the
decontamination factor of Yb in the Es source which was determined to be about
1200 (see Fig. 2 (top)).} it was estimated that the probabilities that
all these isotopes, except $^{154}$Yb and $^{156}$Lu, will be produced
and present in our sources are extremely small, factor of 10$^{5}$ to
10$^{16}$ below the experimental values. (In addition, there are
other arguments which prove that some of these isotopes are not
those present in our spectra, such as the missing of competing
$\alpha$-particle groups or characteristic $\alpha$-groups of
the daughters decays. For instance, the 5.27 MeV group seen in the
Es source can not be due to $^{174m}$Ir, since one does not see
in fig. 4 its 7.3 times stronger $\alpha$-group at 5.48 MeV).

As far as $^{154}$Yb and $^{156}$Lu are concerned, the 5.27 MeV group
seen in the
Es source is not due to $^{154}$Yb, for the following reasons: a) it is
not seen in the Lr-No source, where Yb should be present with about 400
times stronger intensity, and b) one does not see in
Fig. 4 the decay of its descendant,
$^{150}$Dy, at 4.23 MeV
with a total estimated intensity of about 11 counts.
(Only 2 counts, which may  also be
background events, are seen in this energy region. The probability to see
2 events when 11 are expected is 1x10$^{-3}$).
Likewise, the 5.53
MeV group seen in the Lr-No source is not due to $^{156}$Lu,
since one does
not see in Fig. 5  the decay of its descendant, $^{152}$Er, at
4.8 MeV, with 90\% intensity as compared to the 5.53 MeV group.

Thus it is seen that one may rule out the possibility that the experimentally
observed $\alpha$-particle groups are due to unknown
long-lived isomeric states
in the neutron-deficient Yb to Pt nuclei.

\bigskip
Another possibility may perhaps be that the observed $\alpha$-particle
groups are due to unknown long-lived isomeric states in the rare-earth region.
 This possibility is very unlikely
for the following reasons:

a) The decay properties of the rare-earth nuclei were studied
very thoroughly, including measuring of
extremely small branching ratios of $\alpha$-particles, using reactions
which were particularly chosen for the production of each
isotope by itself.
There is no known long-lived isomeric state which decays
by relatively high-energy $\alpha$-particles (5 - 5.5 MeV as compared to
around 3 MeV) in the whole rare-earth region. It is very difficult to see
how such unknown activities will be produced and be present
in our sources,
which were separated
from the rare-earth elements quite thoroughly (the decontamination factor
 of the
Bk and Es sources from their rare-earth homologs was about 10$^{5}$
(see section II above), and that of the Lr-No source
from Yb and Lu was about
4.5x10$^{2}$ (Fig. 2, top and bottom), when they were not seen in the
 experiments in which the rare-earth isotopes were specifically studied.

b) By using the formulas of Ref. 7,
the expected half-lives for such high energy $\alpha$-particles in the
rare-earth region are very short, 2.0x10$^{-2}$ s for 5.14 MeV in
for instance
$^{148}$Eu, 5.6x10$^{-2}$ s for 5.27 MeV in $^{152}$Tb, and 2.4 s for 5.53 MeV
in $^{172}$Lu. (Eu, Tb and Lu are the chemical homologs of Bk,
Es and Lr-No, respectively (Fig. 2)).
 The measured half-lives are longer by factors  of 6x10$^{9}$,
9.6x10$^{8}$ and 9.4x10$^{5}$, respectively. High spin isomers
can explain such large retardations only if $\Delta$L$_\alpha$
is equal to about
15 for the first and second cases, and about 12 for the third case \cite{9}.
Taking into account the large availability of states with various
spins in the
rare-earth nuclei, far from closed shells, and the various
possible decay modes,
it is very difficult to see a
situation where a state with at least such a high spin exists,
and does not have
any other way to decay, but by very long-lived $\alpha$'s
with $\Delta$L$_\alpha$
values of 12 or 15.

c) Another alternative could in principle be the
existence, in rare-earth nuclei, of shape isomeric
states, for
instance long-lived isomeric states in the second  minimum of the
potential in the parent nuclei, which decay, by very retarded
$\alpha$-particles, to the normal states of the daughter.
 (Such states have
been discovered by us \cite{11} in Os, Ir and Hg isotopes using the
reaction between the deformed  $^{28}$Si and  $^{181}$Ta nuclei at
bombarding energy
below and around the Coulomb barrier). However,
no super-deformed minima at low spins have been theoretically found
in rare-earth nuclei with Z$\geq$62 and N${<}$126 \cite{19}.
Superdeformed minima have been produced at high spins
in the A $\sim$ 150 region using heavy-ion reactions.
This region has been studied very carefully. The  minima are at high
energies and the de-excitation is via multi-step or even single step
$\gamma$-ray transitions \cite{20}.
  It is very difficult to see how unknown high-spin isomeric
states will be
produced in the second minima via spallation reactions
 and be present in our sources,
taking into account that high spins are not produced preferentially
by such reactions, and that our sources were separated from nuclei
with A $\sim$ 150, by a factor of about 1x10$^{5}$.

\bigskip
Since it is impossible to identify the observed $\alpha$-particle groups with
any known activity in the whole nuclear chart, and, in addition, as
shown above, they are not due to unknown long-lived isomeric states
in the neutron-deficient Yb to Pt nuclei, nor due to  unknown long-lived
isomeric states in the rare-earth region, one may try to interpret them as
due to some new unknown long-lived isomeric states in the actinide
region. As
 mentioned in the introduction, the production of actinides
 via secondary
reactions in the W target
 have been seen before \cite{2}.
$^{236}$Am and $^{236}$Bk (in
long-lived isomeric states) have been produced and
positively identified by
the decay of their daughter, after $\beta$$^{+}$ or EC decay(s),
$^{236}$Pu
 \cite{2}.

 The measured energies
of the $\alpha$-particle groups
 are low as compared to the
known energies in the heavy actinides \cite{21}, and their
lifetimes are
enhanced by
a factor of 10$^{4}$ to 10$^{7}$ as compared to the systematics \cite{7} of
$\alpha$-particle decay in this region of nuclei.
 From the identification of the isomeric states in the $^{236}$Am and
$^{236}$Bk nuclei \cite{2}, it may be deduced that neutron-deficient
actinide nuclei
are produced via secondary reactions in the W target. Since the intensity
of the 5.14 MeV group in the Bk source grew at the beginning, one may conclude
that this group is emitted from a daughter nucleus, produced by
the $\beta$$^{+}$
or EC process, rather than from an isotope of Bk
itself. (One may exclude $\alpha$-particle decay from the produced Bk isotope
to the daughter nucleus which decays by the 5.14 MeV group, since no group
of $\alpha$-particles was seen in the first measured spectrum, taken about
two and a half years before the one seen in fig.~3 (ref. \cite{2},
fig.~1c).

Tables I and II contain measured and calculated $\alpha$-decay energies
and lifetimes which we use to identify those nuclides emitting these
unidentified $\alpha$-groups. Curium and Americium are the immediate
$\beta^{+}$/EC decay products of Berkelium.
In column 1 of table 1 the $\alpha$-particle energies \cite{21} for
the ground state to ground state transitions of several neutron-deficient Cm
and Am nuclei are given. These energies of 5.6 to 6.5 MeV (some from
experiment and some from systematics), are substantially
higher than the observed  5.14 MeV. It is seen in table 2  that the half-life
of 3.8 y is enhanced by a factor of 1.5x10$^{5}$ or 3.8x10$^{4}$ as compared
to the calculated values \cite{7} for such energy from two typical
parent nuclei,
 $^{238}$Am and $^{238}$Cm, respectively.

It is also shown in column 2 of table 1 that the ground state to ground state
transition energies (mostly from systematics) in neutron-deficient
Es isotopes are in the
range of 7.4 to 8.2 MeV. These energies are much higher than the observed
 5.27 MeV.
Furthermore, as seen in table 2, a half-life of 625 d for such low energy
$\alpha$-particles in Es isotopes is too
fast by about a factor of 2.8x10$^{6}$.

The typical ground state to ground state $\alpha$-energies
for Lr-No isotopes are 8 -- 9 MeV and the
observed 5.53 MeV
is very low compared to these values. (For a reason which will
become clear in sec. III.c, $\alpha$-energies for Fm isotopes
rather than for Lr and No are given in table 1). Also,
a half-life of 26 d for 5.53 MeV
$\alpha$-particles in the Lr-No region, is enhanced by a factor of about
8.5x10$^{6}$ (see table 2).

From this discussion it is obvious that previously unknown isomeric states,
 rather than the normal
ground states, were produced in the reaction. As mentioned in the
introduction, long-lived
isomeric states have recently been observed in the second \cite{9,10,11}
and third \cite{11} wells of the potential energy surfaces.
In what follows,
the unidentified $\alpha$-groups are analyzed in terms of
production of such isomeric states.

In table 1, predicted \cite{12} $\alpha$-particle energies are given for
several neutron-deficient actinide nuclei, for
transitions  from the second minima in the parent nuclei to the ground states
and
to the second and
third minima in the daughters, and from the third minima in the parents to the
third and second minima in the daughters. Table 2 gives calculated \cite{9}
 half-lives
for various transitions from the second and the third minima in the parents,
to the corresponding minima in the daughters. The effect of large deformations
on the $\alpha$-decay was calculated with the potential parameters of
Igo \cite{22} but with a deformed radius\\
R($\theta$) = c($\beta$)R$_{0}$[1 +
$\beta$$_{2}$Y$_{20}$($\theta$) +
$\beta$$_{3}$Y$_{30}$($\theta$) +
$\beta$$_{4}$Y$_{40}$($\theta$)] \hspace{6.0cm} (1)\\
where c($\beta$) is determined from the volume conservation condition and
R$_{0}$ = r$_{0}$ x A$^{1/3}$. A formation
probability for the $\alpha$-particle of \cite{23} 25(4/A)$^{3}$
was used in the calculations.
Only the decay out of the tip ($\theta$ = 0) of the emitting nucleus
was directly calculated. Solid
angle averaging increases the lifetimes by about a factor of 3. A 4\% increase
in the radius parameter (from r$_{0}$ = 1.17 to r$_{0}$ = 1.22 fm), reasonable
for $\alpha$-decay from a highly excited isomeric state, reduces the lifetimes
by about a factor of 4. $\beta$$_{2}$ and $\beta$$_{4}$ values for the
different potential minima were deduced
from the $\epsilon$$_{2}$ and $\epsilon$$_{4}$ values given
in ref. \cite{12}, using fig.~2 of ref. \cite{24}. For third minima transitions,
where $\epsilon$$_{2}$ and $\epsilon$$_{3}$ values are given in ref. \cite{12},
three calculations were performed. In the first one a $\beta$$_{2}$ value
was obtained as described above and $\beta$$_{3}$ was taken as zero. In the
second calculation $\beta$$_{3}$ was taken equal to $\epsilon$$_{3}$, and
in the third calculation the parameters obtained by \'{C}wiok et al.
\cite{25} for the hyperdeformed state in $^{232}$Th of $\beta$$_{2}$ = 0.85,
$\beta$$_{3}$ = 0.35 and $\beta$$_{4}$ = 0.18, were used.

\subsection{The 5.14 MeV $\alpha$-Particle Group in the Bk Source}

It is seen in table 1 that from  the energy point of view the 5.14 MeV
$\alpha$-particle group may correspond
to a II$^{min}$ to II$^{min}$ transition from $^{238}$Am or $^{239}$Am,
or to a III$^{min}$ to III$^{min}$ transition from $^{238}$Cm,
$^{239}$Cm or $^{237}$Am. Table 2 shows that the observed lifetime
is consistent with a II$^{min}$ to II$^{min}$ transition. It is also
consistent with a III$^{min}$ to III$^{min}$ transition, if $\beta$$_{3}$ = 0.
 However, the observed half-life
is retarded by 1 to 3 orders of magnitude if various octupole deformations
are considered. It is seen in fig. 6 that according to the predictions of
Howard and M\"{o}ller \cite{12} the second minimum in $^{238}$Am
is more pronounced than the third minimum in the same nucleus or
in $^{238}$Cm which favors the II$^{min}$ to II$^{min}$ transition.
 In conclusion, the 5.14 MeV group seen in the Bk source may be
explained by an unhindered
 II$^{min}$ to II$^{min}$ transition from $^{238}$Am
or $^{239}$Am. However, the possibility that it is due to a hindered
III$^{min}$ to III$^{min}$ transition from $^{238}$Cm, $^{239}$Cm or
$^{237}$Am can not be ruled out. This is so because
 retardation of $\alpha$-particle decay  may  be due to
small overlap between the initial and final wave functions, even for
transitions between states which belong to the same potential well.
Furthermore, the third
minimum in nuclei around $^{238}$Cm may perhaps be more pronounced than
seen in fig. 6 if
high spins are considered.\footnote{Such a phenomenon is known in the second
well in nuclei like for instance $^{236}$Pu or $^{242}$Cm where two
shape isomeric states
have been observed, and in fact the higher excited state lives longer than the
lower one. (D. N. Poenaru, M. S. Iva\c{s}cu, and D. Mazilu in {\it Particle
Emission from Nuclei}, Eds. D. N. Poenaru and M. S. Iva\c{s}cu, CRC Press,
Vol. III, p. 41 (1989).}

\subsection{The 5.27 MeV $\alpha$-Particle Group in the Es Source}

Let us now discuss the 5.27 MeV group seen in the Es source. As seen in
table 1, columns 6 and 7, it is consistent, from the energy point of view,
with a III$^{min}$ to III$^{min}$ transition or III$^{min}$ to II$^{min}$
transition, from $^{246}$Es or $^{247}$Es. It is seen in table 2 that the
measured half-life is enhanced by about a factor of 70 as compared to
the calculated value, for a II$^{min}$ to II$^{min}$ transition.
 In a III$^{min}$ to II$^{min}$ transition the $\alpha$-particle
sees the potential of the daughter nucleus in the second minimum.
Because of the change in the shape in a transition from the
III$^{min}$ to the II$^{min}$, such a transition will be hindered as compared
to a II$^{min}$ to II$^{min}$ transition. (In the case of the
8.6 MeV III$^{min}$ to II$^{min}$ $\alpha$-transition from
$^{195}$Hg \cite{11}, the hindrance factor as compared to a
II$^{min}$ to II$^{min}$ transition is in the range of 10$^{16}$).
 This means that the enhancement
factor for the lifetime of the 5.27 MeV assuming III$^{min}$ to II$^{min}$
transition, should be much larger than 70. But it is
quite difficult to understand
such a large enhancement and it is much more reasonable to assume
that this group
is due to a III$^{min}$ to III$^{min}$ transition.
Considering a
III$^{min}$ to III$^{min}$ transition, it is seen in table 2 that the
measured
half-life is enhanced by a factor of 20 if $\beta$$_{3}$ = 0, and it is
retarded by a factor of 1.6 -- 120 if $\beta$$_{3}$ and also $\beta$$_{4}$
are included in the calculations. Since it is easier to understand
retardation of  $\alpha$-decay rather than enhancement, in transitions
within the same potential well, it seems that a quadrupole deformation
 alone is not
enough to describe the hyperdeformed states in the daughter nuclei around
$^{243}$Bk.

Fig. 6 shows \cite{12} that in the second minimum of $^{247}$Es the outer
barrier is quite low, while for the third minimum the inner barrier is very
large, and in fact the third minimum is predicted to be the true ground
state of this nucleus, being 0.61 MeV below the normal, low deformed,
ground state \cite{12}. Unfortunately there are no predictions in this case
for the
potential at even larger deformations, beyond the third minimum. (In the case
of $^{232}$Th \cite{25} the outer barrier in the third minimum is quite high).

\subsection{The 5.53 MeV $\alpha$-Particle Group in the Lr-No Source}

To the best of our knowledge there are no predictions for the second and third
minima for Lr-No nuclei. Therefore the predictions \cite{12} for
 neutron-deficient Fm isotopes were used in deducing the various
$\alpha$-energies given in table 1. It is seen that such a low
energy as 5.53 MeV may only be due to an isomeric state in the third well
of the potential. From the energy point of view it may be either due to a
III$^{min}$ to III$^{min}$ or a III$^{min}$ to II$^{min}$ transition.
By extrapolating according to the $\alpha$-energies from $^{246}$Es
and $^{248}$Fm,
one finds the parent isotope  around $^{252}$No or $^{254}$Lr.
It is seen in table 2 that the measured half-life of 26 days is enhanced
by a factor of about 160 as compared to the prediction assuming a
II$^{min}$ to II$^{min}$ transition. Based on the arguments given above one
may conclude that the 5.53 MeV group is due to a III$^{min}$ to III$^{min}$
and not due to a III$^{min}$ to II$^{min}$ transition.
Regarding a III$^{min}$ to III$^{min}$ transition, it is seen in table 2
that the measured half-life is still enhanced by a factor of about 85 if only
$\beta$$_{2}$ deformation is taken into account in the calculations. However,
a consistency is obtained when $\beta$$_{3}$ or $\beta$$_{3}$ plus
$\beta$$_{4}$ are included.
Fig. 6 shows that the features mentioned above in $^{247}$Es occur also in
$^{248}$Fm. The outer barrier in the second minimum is low and the inner
barrier in the third minimum is high, and the third
minimum is in fact predicted \cite{12} to be the true ground state of this
nucleus, being 1.76 MeV below the ground state of the first minimum.
One may assume that the potential-energy curves of the nuclei around
 $^{252}$No -- $^{254}$Lr are similar, explaining the observation of the
5.53 MeV group in the Lr-No source.\footnote{It should be mentioned that
in addition to various theoretical predictions about the existence of the
third minimum in the potential energy in various actinide nuclei which include
refs. [12] and [19] and which was first proposed by
 P. M\"{o}ller and J.R. Nix
(Proc. Int. Conf. on Physics and Chemistry of Fission, Rochester, 1073 (IAEA,
Vienna 1974) Vol. I, p. 103), experimental evidences were also obtained in
Th and U nuclei by
 J. Blons et al. (Nucl. Phys. {\bf A477}, 231 (1988)), and
 A. Krasznahorkay et al. (APH N.S., Heavy
Ion Physics {\bf 7}, 35 (1998)).}

Table 3  summarizes the half-lives of the isomeric states as compared to the
half-lives of their corresponding normal-deformed ground states. It is
assumed in this comparison that the 5.14, 5.27 and 5.53 MeV groups are from
$^{238}$Am, $^{247}$Es and $^{252}$No nuclei, respectively (see table 1).
For completeness the data on the isomeric states in $^{236}$Am and $^{236}$Bk
\cite{2} are also included in this table. It is seen that the lifetimes of
the isomeric states are 10$^{4}$ to 10$^{6}$ longer as compared to their
corresponding ground states of normal deformations.

\subsection{Production Cross Sections of the Various Isotopes}

As mentioned in the introduction the production yield in the
secondary reactions depends on two cross sections: a) the cross
section for production of suitable fragments with high enough kinetic energies,
which may interact with other W nuclei in the target, and
 b) the heavy-ion reaction cross section between such
fragments and other W nuclei in the target. In the following we will try to
deduce the cross sections for the second reaction, based on reasonable
estimates about the first
step of the reaction, the production of the high-energy fragment.

Table 4 summarizes the process of deducing the heavy-ion reaction
cross sections. The data regarding the production of the
isomeric states in $^{236}$Am and $^{236}$Bk \cite{2} are also included
in this table. The numbers of produced atoms, which were deduced from the
measured intensities and half-lives of the $\alpha$-lines, are given in column
2. One can see that 3x10$^{5}$ to 1.4x10$^{4}$ atoms were produced in the
isomeric states of the various isotopes. Column 3 gives the total cross
sections which are in the 10$^{-12}$ to 10$^{-14}$ b region.
These cross sections
were deduced from the number of produced nuclei, taking into account the
integrated number of protons in the beam
  and the thickness of the target. In column 4 the
most suitable possible reactions are listed assuming that the bombarding
energies of the relevant secondary fragments   are
around the  Coulomb barriers between
 the projectile and
target nuclei, at 4.2 to 4.8 MeV/nucleon. It is seen in column 5 that
in general, at E$_{Lab}$ equal to the Coulomb barrier, the reactions
are quite hot
with excitation energies of the
compound nuclei around 30 -- 40 MeV.
 The yield from higher bombarding
energies is presumably smaller since first, the production of high energy
fragments is going down as their outgoing energy increases \cite{17},
 and secondly, producing the
compound nucleus at higher excitation energy results in a lower evaporation
residue cross section, because of the competition with fission during
the cooling process. Only reactions with evaporation of a proton,
an $\alpha$-particle, or a
proton + a neutron and an $\alpha$-particle + a neutron were considered.
(The yields from the (HI,xn) reactions, in particular from the (HI,3n)
 reactions, are expected to be less important: first,
higher neutron-excess fragments which act as projectiles are needed, and these
 are produced less easily;
 and secondly,
one loses more yield, due to competition with fission, in three
step evaporation processes as
compared to one or two
steps\footnote{For instance, at E$_{Lab}$ = 154 MeV (E$_{x}$(C.N.) = 51.9 MeV),
cross sections of about
70 and 37 $\mu$b were obtained for the $^{175}$Lu($^{32}$S,p2n)$^{204}$Rn
and the $^{175}$Lu($^{32}$S,$\alpha$2n)$^{201}$At reactions respectively,
as compared to about 7 $\mu$b for the $^{175}$Lu($^{32}$S,4n)$^{203}$Fr and
$^{175}$Lu($^{32}$S,3n)$^{204}$Fr reactions
(S. Gelberg, Ph. D. thesis, 1991).}).
The same cross sections were assumed for the various heavy-ion
reactions which contribute to the production of a particular isotope.
The range of the total
production cross sections of the various fragments listed in column 4, are
given in column 6. These cross sections were estimated using first the
measured cross sections for the production of various Sc isotopes in the
interaction of 18.2 GeV protons on a Ta target \cite{26}. (These values are
very similar to
 those obtained in the interaction of 28.5 GeV protons on Pb \cite{27}).
Secondly,
 by using exponential extrapolation, values for heavier Sc
isotopes
were obtained. The cross sections for the various fragments considered in
table 4 were then obtained using A/Z scaling. Column 7 in table 4 gives the
values of "f", the ratio of the production cross sections for high energy
fragments, above or near the Coulomb barrier, to the total cross sections of
 the corresponding
 fragments. It was assumed to be 1\% for the fragments which
produced the $^{236}$Am isotope, and reduced exponentially to 0.1\% for those
which produced the $^{252}$No isotope. (A cross section of about 20 $\mu$b
is deduced for Ar ions of $\geq$4.5 MeV/nucleon produced in the interaction of
5.5-GeV protons with U (fig. 14 of ref. \cite{17}). Therefore,
cross sections of about 8 $\mu$b
or 0.9 $\mu$b for the production of high energy $^{51}$Ti
or $^{67}$Cu ions respectively, in the interaction of 24-GeV protons
with W, are
reasonable).

The last column in table 4 gives the deduced evaporation residue cross
sections  which
are from 0.2 to 13 mb. The results depend somewhat on the
assumptions made above. However,
it seems that first, the cross sections do not decrease
when the Z-value of the evaporated residual nucleus increases from 95 to 102,
and secondly, the cross sections are very large as compared to
known heavy-ion reactions, in particular when $^{252}$No is considered,
where the typical cross sections are below 1~$\mu$b \cite{8}.

Fig.~7 shows calculated fusion cross sections using the coupled-channel
deformed fusion code
of Ref. \cite{28} for a typical $^{70}$Zn + $^{186}$W reaction.
Curve 7a was obtained
 assuming that the projectile and the target
nuclei are spherical. In curve 7b the known \cite{29} quadrupole
deformation
of the target nucleus was included in the calculations. Curve 7c
 displays the
results when coupling to the 2$^{+}$ and 3$^{-}$ vibrational states
of the
projectile nucleus were also taken into account. In the secondary
reaction experiments, the projectile is a fragment which had been produced
only within about 2x10$^{-14}$ s before interacting with another W nucleus in
the target. During this short time it is still at high excitation energy
and quite deformed. In particular, since it is produced as a deformed
fragment, it may have a deformation which is typical to the second
or third
well of the potential, rather than to normal deformation. Curve 7d
shows the results of calculations taking into account the known
quadrupole deformation of the target nucleus, and assuming for the projectile
fragment ($^{70}$Zn) a $\beta$$_{2}$ value of 0.6, a somewhat higher value than
typical for the second minimum of
the potential in this region of nuclei \cite{30}.
According to  curve 7a,  a cross section of about 10 mb is predicted
 at bombarding energies around the Coulomb barrier
of 232 MeV.
This result can not explain the experimental values, since one would fall
considerably below the measured evaporation residue cross sections,
allowing for competition with fission in producing the final evaporation
residual nuclei. (One gets larger fusion cross sections at higher
bombarding energies, but than the number of evaporated particles
increases, which will decrease the evaporation residue
production cross section).
 In curves 7b and 7c,  fusion cross sections of
60 -- 70 mb are
predicted at bombarding energies around the Coulomb barrier.
An even larger fusion cross section of about 160 mb
is predicted (curve 7d)
when a large deformation  for the fragment projectile is assumed.
These predictions
are consistent with the experimental results
if $\Gamma$$_{n}$/$\Gamma$$_{f}$,
$\Gamma$$_{p}$/$\Gamma$$_{f}$ and $\Gamma$$_{\alpha}$/$\Gamma$$_{f}$
are not
too small. It is also seen in curve 7d that the very cold fusion reaction,
i.e. the radiative capture reaction at E$_{c.m.}$ $\sim$ 215 MeV,
below the one
neutron separation energy where the competition with fission is
very much
reduced, is perhaps possible with a fusion cross section around 30 mb. (If one
would have assumed only this reaction as compared to the four reactions
(p, pn, $\alpha$ and $\alpha$n) considered in table 4,
a four times larger cross sections than those
given in the last column of table 4 are needed in order to be consistent
with the experimental results).
Thus, the
unusually large evaporation residue cross sections deduced from
the secondary reaction
experiments, can consistently be understood, provided that very little
 extra-push
energy \cite {31,32,33} is needed in order to produce the compound
nucleus.
 An
extra-push
energy means that one has to shift the plots of fig. 7 to the right
by its
value. It is seen in curves 7b and 7c that around 232 MeV, an extra-push
energy of 5 MeV reduces the fusion cross sections down to 25 -- 35 mb.
This is
pretty low, remembering that one or two particles have to be
evaporated,
and still leave around 10~mb for the production cross section of the
residual nucleus. For curve 7d zero extra-push energy
is consistent with the data
for bombarding energies around 215 MeV, while around 232 MeV,
an extra-push
energy around 10 MeV is tolerable. It should be mentioned that extra-push
energies of about 300 \cite{31}, 50 \cite{32} and 20 MeV \cite{33} were
predicted for the $^{70}$Zn + $^{186}$W reaction. The small
 experimentally deduced extra-push values,
 of 0 -- 10 MeV, are consistent with the previous conclusion that
the compound nucleus was formed in the second or third well of the potential,
and not in the innermost well.

Fig. 8 shows several calculated shapes of the compound nuclei, using Eq. (1)
 for
the deformed radii  and various values for the deformation
parameters. The calculated
shapes of the  target and corresponding projectile nuclei are also
given in the figure. In fig. 8 (top) two calculated
shapes for the compound nucleus with A = 239 are displayed. The first
one (left)
shows the calculated shape of the compound nucleus with deformation
parameters \cite{12} of the normal ground state, while in the second one
(right) the deformation parameters \cite{12} of the second minimum
were used in the calculations. Three calculated shapes for A = 253
are seen in fig. 8 (bottom). In the left side the parameters of the
normal ground state were used in the calculation, while in the middle
only the parameter $\beta$$_{2}$ of ref. \cite{12}
 for the third minimum of the potential
was used.
 In the right (bottom) side of fig. 8 the deformation
parameters
 for the hyperdeformed minimum in $^{232}$Th \cite{25} were taken into
account in our calculations. It is qualitatively seen that much less
rearrangement and matter penetration accompanied by dissipation, and hence
much less
extra-push energy, is needed to form the compound nucleus
in the second and particularly in the third
well of the potential, compared to formation in the
normal states.

\subsection{Consequences Regarding the Production of the Long-Lived
Superheavy Element with Z = 112}

The existence of the long-lived isomeric states in the second and third wells
of the potential can explain, in a very consistent manner \cite{34},
the discovery
\cite{3,4,5} of the superheavy element with Z = 112. The
experimental evidence for
the existence of this element was based mainly on the observation of fission
fragments in Hg sources which were separated from the CERN W target \cite{3},
 and on
the mass measurements of the fissioning nuclei, which were interpreted as due
to five different molecules of the isotope with Z = 112 and N $\cong$ 160
\cite{5}. While it was impossible to disprove the data it seemed very
difficult
to understand them \cite{35}, in particular the long lifetime of the observed
activities of several weeks, and the large deduced heavy ion reaction
 cross section of a
few  mb.
It was pointed out \cite{5} that the radiative capture processes like
$^{184}$W($^{88}$Sr,$\gamma$)$^{272}$112 or
$^{186}$W($^{86}$Sr,$\gamma$)$^{272}$112 are possible at bombarding
energies around the Coulomb barrier. Such reactions in principle could yield
large evaporation residue cross sections. However, it was thought that
the large
extra-push energy expected for such reactions will drastically reduce
the fusion cross section.

The existence of the long-lived isomeric states in
the second and third wells of the potential, and the relatively very little
extra-push energy required for their production, provide  a consistent
interpretation
to the experimental observations. The long lifetime of several weeks
shows that the $^{\sim272}$112 nucleus was probably produced in a long-lived
isomeric state rather than in the normal ground state where
a half-life of about 240 $\mu$s was observed for the $^{277}$112 isotope
\cite{36}. The large deduced fusion cross section of a few mb
(\cite{5} and the discussion
 below), as compared to the evaporation residue cross section of
 about 1 pb for the production of $^{277}$112
\cite{36}, strongly indicates that the isomeric state is in the second or
third well of the potential,
rather than in the first normal-deformed well. (The observation of
the isomeric states in the third well, which may actually be the true ground
states in the very heavy Es and Lr-No nuclei, may imply  a similar
situation also  in the superheavy nuclei region around Z = 112).
 In table 4  a cross section of 3.8 mb
 for the second step of the reaction is deduced from the data,
 assuming a contribution from the radiative capture
process alone. (Adding contributions from other reactions like ($^{x}$Sr,n),
($^{x}$Y,p) or ($^{x}$Zr,$\alpha$), will decrease the deduced cross section).
This value is about the same as those obtained in the actinides. However,
in the actinides, relatively hot reactions took place, while in the Sr + W
reactions an extremely cold type of reaction, the radiative capture
 with excitation
energies of the compound nuclei of 2.0 - 10.0 MeV (table 4), is possible.
The lower fusion cross section in the superheavy element
region of about 4 mb as compared to the predicted 60 -- 160  mb in the
actinides
(see above), may be due to a larger extra-push energy
needed in the superheavies as compared to the actinides. It was shown that
extra-push energies around 30 -- 40 MeV are consistent with the measured
fusion cross section of about 4 mb, assuming various deformations (figs.~4
and 5 of ref. \cite{34}), or deformations which are typical to the
second minimum of the
potential (fig.~3 of ref. \cite{37}), for the secondary Sr fragment.
An extra-push energy of 30 -- 40 MeV
is small compared to the predictions of $\gg$300 \cite{31} and $\gg$50
\cite{32} made for the production of the compound
nucleus in the normal states. (It is compatible with an extra-push
value of $\sim$40 MeV predicted for this reaction in Ref. \cite{33}).
 It is in accord with the
values of 0 -- 10 MeV obtained above in the actinides for producing the
compound nucleus in the second or third well of the potential-energy
surfaces. Thus, the combined effect of first, producing the compound nucleus
in the third or second well of the potential, which requires much lower
extra-push energy as compared to its production in the normal states,
and secondly, the reduced
Coulomb repulsion between a W target nucleus and a deformed fragment
(produced in the
secondary reaction within about 2x10$^{-14}$ s before interacting with
the W target), as compared to
the Coulomb repulsion
caused by the interaction with a normal projectile,
enables one to understand the large
fusion cross section, even if an extra-push energy
of 30 -- 40 MeV were required in the process.

\section{Summary}

The problems of the low energies, large lifetime
enhancements and high production cross sections of the
$\alpha$-particle groups of 5.14, 5.27 and 5.53 MeV, seen
in the Bk, Es and Lr-No sources, respectively, have consistently been
interpreted as due to
formation of long-lived isomeric states
 in the second and third wells of the
potential-energy surfaces. The low energies fit well with theoretical
predictions \cite{12} for II$^{min}$ to II$^{min}$ (or III$^{min}$
to III$^{min}$) transition in the first case, and to III$^{min}$ to
III$^{min}$
transitions in the second and third cases.
The large enhancements of
the $\alpha$-particle decays are due to the increased penetrabilities through
the barriers of superdeformed or hyperdeformed nuclei. The large
production cross sections are due to two effects: a) the rather little,
if any, extra-push energy needed to produce the compound nucleus in
the second or third well of the potential-energy surfaces. b) the
increased fusion cross section due to the deformation of the projectile
(fragment)
in the secondary reactions, as compared to normal projectiles. This
increased fusion cross section can explain the  observed production
of the evaporation residue nuclei, even for an extra-push energy
up to 10 MeV. According to theoretical
predictions \cite{12}, the isomeric states in the third well in
the Es and Lr-No cases, may actually be the true
(or nearby the true) ground states of
the particular nuclei.

The existence of long-lived isomeric states in the second
and third wells of the
potential-energy surfaces seems to be a general phenomenon in
heavy nuclei. They have been seen in neutron-deficient nuclei from the
Os -- Hg region \cite{11}, through Fr \cite{9,10}, and up to the heavy actinide
nuclei,
like Am, Bk, Es and Lr-No. In addition to their very long lifetimes,
much longer than that of their corresponding normal-deformed ground states,
they may decay by very enhanced
$\alpha$-particle emission, in II$^{min}$ to II$^{min}$ \cite{9} or
III$^{min}$ to III$^{min}$ (the present work) transitions, or by very retarded
$\alpha$-particle decay, in II$^{min}$ to I$^{min}$ or in
III$^{min}$ to II$^{min}$
transitions \cite{11}, and also by long-lived proton radioactivity \cite{10}.

It was shown that the existence of the isomeric states in the second and third
well of the potential-energy surfaces enables one to understand, in a fully
consistent manner, the production of the long-lived superheavy element with
Z = 112 \cite{3,4,5,34}, much longer than seen in Ref. \cite{36}.
 In general the discovery of such isomeric states
with their very unusual decay properties, the very much reduced extra-push
energy needed for their production in interactions between very heavy ions,
and the possibility that the third minimum is actually the true ground state
of very heavy and perhaps superheavy nuclei, provide completely new
considerations in the study of heavy and superheavy elements.

\section{Acknowledgements}

 We are grateful to C.~J.~Batty,
A.~I.~ Kilvington, G.~W.~A.~Newton,
J.~D.~Hemingway and S.~Eshhar for taking part in the first stages of the
experiments. The late A.~M.~Friedman and V.~J.~Robinson also
participated at the beginning of the
experiments. We
 appreciate very much the valuable discussions with
 N.~Zeldes and the technical assistance  of S.~Gorni,
O.~Skala and the electronic team of the Hebrew University.
  D.~K. acknowledges the financial support of the DFG.

\newpage
\begin{table}[h]
\caption {$\alpha$-particle transition energies for various
actinide isotopes. Ground state to ground state transition
energies were taken from Audi et al. [21]. Calculated values for
transitions from the second minima of the potential in the parent
nuclei to the ground states and to the second and third minima in
the daughters, and from the third minima in the parents to the
third and second minima in the daughters, were taken from the
predictions of Howard and M\"{o}ller [12]. Numbers in {\bf bold}
are consistent with the experimental results (see text).}
\begin{center}
\begin{tabular}{lllllll}
Mother & E$_\alpha$   & E$_\alpha$ &  E$_\alpha$ & E$_\alpha$ &
E$_\alpha$ & E$_\alpha$  \\ Isotope & g.s.$\rightarrow$g.s. &
II$^{min}$$\rightarrow$II$^{min}$ & II$^{min}$$\rightarrow$g.s. &
II$^{min}$$\rightarrow$III$^{min}$ &
III$^{min}$$\rightarrow$III$^{min}$ &
III$^{min}$$\rightarrow$II$^{min}$\\ \hline $^{238}$Cm & 6.51 &
5.90 & 8.25 &  4.38 & $\bf{5.24}$ & 6.76 \\ $^{239}$Cm & 6.47 &
5.61 & 7.93 &  3.71 & $\bf{4.93}$ & 6.72 \\ $^{240}$Cm & 6.29 &
5.68 & 7.80 &  3.65 & 4.57 & 6.61 \\ $^{241}$Cm & 6.08 & 5.94 &
7.71 &  3.59 & 4.33 & 6.67 \\ \hline $^{237}$Am & 6.08 & 5.50 &
8.21 &  3.58 & $\bf{4.90}$ & 6.88 \\ $^{238}$Am & 5.94 &
$\bf{5.13}$ & 7.76 & 2.88 & 4.53 & 6.78 \\ $^{239}$Am & 5.82 &
$\bf{5.24}$ & 7.67 & 2.78 & 4.16 & 6.72 \\ $^{240}$Am & 5.61 &
5.53 & 7.55 & 2.70 & 3.96 & 6.23\\ \hline
$^{241}$Es & 8.18 & 7.45 & 8.89 & 7.20 & 6.39 & 6.65 \\ $^{242}$Es
& 8.09 & 7.31 & 8.79 & 6.98 & 6.20 & 6.52 \\ $^{243}$Es & 7.94 &
7.27 & 8.82 & 7.03 & 6.02 & 6.26 \\ $^{244}$Es & 7.90 & 7.61 &
9.00 & 7.22 & 5.96 & 6.35 \\ $^{245}$Es & 7.78 & 7.72 & 9.20 &
7.53 & 5.81 & 6.01 \\ $^{246}$Es & 7.61 & 7.83 & 9.44 & 7.81
&$\bf{5.62}$ & $\bf{5.64}$ \\ $^{247}$Es & 7.37 & 7.47 & 9.41 &
7.91 & $\bf{5.27}$ & $\bf{4.83}$ \\ \hline $^{242}$Fm & 8.31$^{a}$
& 7.53 & 8.39 & 7.67 & 6.42 & 6.22 \\ $^{243}$Fm & 8.55 & 7.74 &
8.66 & 7.85 & 6.53 & 6.48 \\ $^{244}$Fm & 8.41 & 7.71 & 8.73 &
8.01 & 6.43 & 6.19 \\ $^{245}$Fm & 8.30 & 7.95 & 8.86 & 8.08 &
6.28 & 6.15 \\ $^{246}$Fm & 8.24 & 8.10 & 9.12 & 8.42 & 6.13 &
$\bf{5.80}$ \\ $^{247}$Fm & 8.06 & 8.18 & 9.39 & 8.73 & 5.90 &
$\bf{5.35}$ \\ $^{248}$Fm & 7.87 & 7.85 & 9.39 & 8.47 &
$\bf{5.62}$ & 4.60
\end{tabular}
\end{center}
\end{table}
\vspace{-1.2cm} \noindent$^{a}$Taken from S.~Liran and N.~Zeldes,
{\it At. Data Nucl. Data Tables} {\bf 17}, 431 (1976).\\




\newpage
\begin{table}[h]
\caption {Predicted half-lives for the 5.14, 5.27 and 5.53 MeV
$\alpha$-particle groups seen in the Bk, Es and No-Lr sources
respectively, assuming transitions from nuclei which are suggested
in Table 1.
 Calculated values for transitions from the second minima of
the potential in the parent nuclei to the second minima in the
daughters, and from the third minima in the parents to the third
minima in the daughters,
 are given ([9] and section III of the text).
The observed half-lives are also compared to those calculated by
the formulas of Viola and Seaborg [7].}
\begin{center}
\begin{tabular}{llllll}
Transition  & $\beta$$_{2}$   & $\beta$$_{3}$ & $\beta$$_{4}$ &
t$_{1/2}^{cal}$ (s)$^{a}$ & t$_{1/2}^{cal}$/ t$_{1/2}^{exp}$ \\
\hline \multicolumn{6}{l} {E$_{\alpha}$ = 5.14 MeV;
t$_{1/2}^{exp}$ = 3.8$\pm{1}$ y = 1.2x10$^{8}$ s}\\
\multicolumn{6}{l}{$^{238}$Am$\rightarrow$$^{234}$Np;
t$_{1/2}$(V.S.)$^{b}$ = 1.8x10$^{13}$ s;
t$_{1/2}^{V.S.}$/t$_{1/2}^{exp}$ = 1.5x10$^{5}$ }\\ \hline
II$^{min}$$\rightarrow$II$^{min}$ & 0.71$^{c}$ & 0.0 & 0.09$^{c}$
& 3.4x10$^{8}$ & 2.8 \\ III$^{min}$$\rightarrow$III$^{min}$ &
1.05$^{c}$ & 0.0$^{d}$ & 0.0 & 1.0x10$^{8}$ & 8.3x10$^{-1}$ \\
III$^{min}$$\rightarrow$III$^{min}$ & 1.05$^{c}$ & 0.18$^{e}$ &
0.0 & 4.5x10$^{6}$ & 3.8x10$^{-2}$ \\
III$^{min}$$\rightarrow$III$^{min}$ & 0.85$^{f}$ & 0.35$^{f}$ &
0.18$^{f}$ & 5.1x10$^{4}$ & 4.3x10$^{-4}$ \\ \hline \hline
\multicolumn{6}{l} {E$_{\alpha}$ = 5.14 MeV; t$_{1/2}^{exp}$ =
3.8$\pm{1}$ y = 1.2x10$^{8}$ s}\\
\multicolumn{6}{l}{$^{238}$Cm$\rightarrow$$^{234}$Pu;
t$_{1/2}$(V.S.)$^{b}$ = 4.5x10$^{12}$ s;
t$_{1/2}^{V.S.}$/t$_{1/2}^{exp}$ = 3.8x10$^{4}$ }\\ \hline
III$^{min}$$\rightarrow$III$^{min}$ & 1.05$^{c}$ & 0.0$^{d}$ & 0.0
& 3.3x10$^{8}$ & 2.7 \\ III$^{min}$$\rightarrow$III$^{min}$ &
1.05$^{c}$ & 0.17$^{e}$ & 0.0 & 1.7x10$^{7}$ & 1.4x10$^{-1}$ \\
III$^{min}$$\rightarrow$III$^{min}$ & 0.85$^{f}$ & 0.35$^{f}$ &
0.18$^{f}$ & 1.5x10$^{5}$ & 1.2x10$^{-3}$ \\ \hline \hline
\multicolumn{6}{l} {E$_{\alpha}$ = 5.27 MeV; t$_{1/2}^{exp}$ =
625$\pm{84}$ d = 5.4x10$^{7}$ s}\\
\multicolumn{6}{l}{$^{247}$Es$\rightarrow$$^{243}$Bk;
t$_{1/2}$(V.S.)$^{b}$ = 1.5x10$^{14}$ s;
t$_{1/2}^{V.S.}$/t$_{1/2}^{exp}$ = 2.8x10$^{6}$ }\\ \hline
II$^{min}$$\rightarrow$II$^{min}$ & 0.71$^{c}$ & 0.0 & 0.09$^{c}$
& 3.9x10$^{9}$ & 7.2x10$^{1}$ \\
III$^{min}$$\rightarrow$III$^{min}$ & 1.05$^{c}$ & 0.0$^{d}$ & 0.0
& 1.1x10$^{9}$ & 2.0x10$^{1}$  \\
III$^{min}$$\rightarrow$III$^{min}$ & 1.05$^{c}$ & 0.19$^{e}$ &
0.0 & 3.3x10$^{7}$ & 6.1x10$^{-1}$ \\
III$^{min}$$\rightarrow$III$^{min}$ & 0.85$^{f}$ & 0.35$^{f}$ &
0.18$^{f}$ & 4.6x10$^{5}$ & 8.5x10$^{-3}$ \\ \hline \hline
\multicolumn{6}{l} {E$_{\alpha}$ = 5.53 MeV; t$_{1/2}^{exp}$ =
26$\pm{7}$ d = 2.2x10$^{6}$ s}\\
\multicolumn{6}{l}{$^{252}$No$\rightarrow$$^{248}$Fm;
t$_{1/2}$(V.S.)$^{b}$ = 1.9x10$^{13}$ s;
t$_{1/2}^{V.S.}$/t$_{1/2}^{exp}$ = 8.5x10$^{6}$ }\\ \hline
II$^{min}$$\rightarrow$II$^{min}$ & 0.79$^{c}$ & 0.0 & 0.14$^{c}$
& 3.6x10$^{8}$ & 1.6x10$^{2}$ \\
III$^{min}$$\rightarrow$III$^{min}$ & 1.2$^{c}$ & 0.0$^{d}$ & 0.0
& 1.9x10$^{8}$ & 8.5x10$^{1}$ \\
III$^{min}$$\rightarrow$III$^{min}$ & 1.2$^{c}$ & 0.19$^{e}$ & 0.0
& 7.0x10$^{6}$ & 3.1 \\ III$^{min}$$\rightarrow$III$^{min}$ &
0.85$^{f}$ & 0.35$^{f}$ & 0.18$^{f}$ & 2.3x10$^{5}$ &
1.0x10$^{-1}$ \\
\end{tabular}
\end{center}
\end{table}
\vspace{-0.9cm} \noindent$^{a}$ See text.\\ $^{b}$ Ref. [7].\\
$^{c}$ $\epsilon_{2}$ and $\epsilon_{4}$ were taken from Ref. [12]
 and converted to $\beta_{2}$ and $\beta_{4}$ according to Ref. [24].\\
$^{d}$ Assuming $\beta_{3}$ = 0.\\ $^{e}$ Assuming $\beta_{3}$ =
$\epsilon_{3}$ of Ref. [12].\\ $^{f}$ Parameter given by \'{C}wiok
et al. [25] for the hyperdeformed state of $^{232}$Th.\\
\newpage
\begin{table}[h]
\caption {Half-lives of some isomeric states and their ratios to
the half-lives of their corresponding normal-deformed ground
states. }
\begin{center}
\begin{tabular}{llll}

Isotope & t$_{1/2}$ (g.s.) & t$_{1/2}$ (i.s.) & t$_{1/2}$
(i.s.)/t$_{1/2}$ (g.s.)\\ \hline
$^{236}$Bk &  42.4 s$^{a}$ & $\geq$ 30 d$^{b}$ &  $\geq$ 6.1 x
10$^{4}$\\ $^{236}$Am &  3.7 m(?)$^{c}$ & 219 d$^{b}$  &   8.5 x
10$^{4}$\\ $^{238}$Am$^{d}$ & 98 m$^{g}$  & 3.8 y & 2.0 x
10$^{4}$\\ $^{247}$Es$^{e}$ & 4.55 m$^{g}$ & 625 d &  2.0 x
10$^{5}$\\ $^{252}$No$^{f}$ & 2.3 s$^{g}$ & 26 d &  9.8 x
10$^{5}$\\
\end{tabular}
\end{center}
\end{table}
\vspace{-0.9cm} \noindent$^{a}$ Predicted by P.~M\"{o}ller, J.
R.~Nix and K.-L.~Kratz, {\it At. Data Nucl. Data Tables} {\bf 66},
131 (1997).\\ $^{b}$ Ref. [2].\\ $^{c}$ G.~Pfennig,
H.~Klewe-Nebenius and W.~Seelmann-Eggebert, Karlsruher Nuklidkarte
(1995).\\ $^{d}$ Assuming that the 5.14 MeV is from $^{238}$Am
(see table 1).\\ $^{e}$ Assuming that the 5.27 MeV is from
$^{247}$Es (see table 1).\\ $^{f}$ Assuming that the 5.53 MeV is
from $^{252}$No (see text).\\ $^{g}$ R.~B.~Firestone et al., {\it
Table of Isotopes}, Wiley-Interscience (1996).\\




\newpage
\begin{table}[h]
\caption {Estimated production cross sections of various actinide
nuclei in their isomeric states via secondary reactions.
Estimation for the superheavy isotope $^{272}$112 is also given.}
\begin{center}
\begin{tabular}{llllllll}
Isotope& No. of  & $\sigma$$_{T}$ & Reactions &
E$_{x}$(C.N.)$^{a}$ & $\sigma$$_{frag.}$$^{b}$ & f$^{c}$ &
$\sigma$$_{R}$(H.I.)$^{d}$  \\
   & Atoms  & (barn)  &   &
(MeV) & (mb) &  & (mb) \\ \hline $^{236}$Am & 3.1x10$^{5}$ &
8.2x10$^{-13}$ & $^{182-184,186}$W($^{55-53,51}$Ti,p) & 29.8--38.0
& 0.77--0.01 & 1x10$^{-2}$ & 12.8 \\
  &  &  &  $^{182-184,186}$W($^{56-54,52}$Ti,pn) &  &  &  &  \\
  &  &  &  $^{182-184,186}$W($^{58-56,54}$V,$\alpha$) &  &  &  &  \\
  &  &  &  $^{182-184,186}$W($^{59-57,55}$V,$\alpha$n) &  &  &  &  \\
$^{236}$Bk & 4.4x10$^{4}$ & 1.2x10$^{-13}$ &
$^{182-184,186}$W($^{55-53,51}$Cr,p) & 30.9--42.2 & 5.5--0.4  &
5x10$^{-3}$ & 0.22 \\
  &  &  & $^{182-184,186}$W($^{56-54,52}$Cr,pn)  &  &  &  &  \\
  &  &  & $^{182-184,186}$W($^{58-56,54}$Mn,$\alpha$)  &  &  &  &  \\
  &  &  & $^{182-184,186}$W($^{59-57,55}$Mn,$\alpha$n)  &  &  &  &  \\
$^{238}$Bk & 3.3x10$^{4}$ & 8.9x10$^{-14}$ &
$^{182-184,186}$W($^{57-55,53}$Cr,p) & 33.1--36.9 & 4.0--0.09 &
5x10$^{-3}$ & 0.26 \\
  &  &  & $^{182-184,186}$W($^{58-56,54}$Cr,pn)  &  &  &  &  \\
  &  &  & $^{182-184,186}$W($^{60-58,56}$Mn,$\alpha$)  &  &  &  &  \\
  &  &  & $^{182-184,186}$W($^{61-59,57}$Mn,$\alpha$n)  &  &  &  &  \\
$^{247}$Es & 1.4x10$^{4}$ & 3.7x10$^{-14}$ &
$^{182-184,186}$W($^{66-64,62}$Fe,p) & 33.3--31.4 & 0.24--0.006 &
2.6x10$^{-3}$ & 6.2 \\
  &  &  & $^{182-184,186}$W($^{67-65,63}$Fe,pn)  &  &  &  &  \\
  &  &  & $^{182-184,186}$W($^{69-67,65}$Co,$\alpha$)  &  &  &  &  \\
  &  &  & $^{182-184,186}$W($^{70-68,66}$Co,$\alpha$n)  &  &  &  &  \\
$^{252}$No & 1.4x10$^{4}$ & 3.7x10$^{-14}$ &
$^{182-184,186}$W($^{71-69,67}$Cu,p) & 24.4--27.4 & 0.88--0.03 &
1x10$^{-3}$ & 4.0 \\
  &  &  & $^{182-184,186}$W($^{72-70,68}$Cu,pn)  &  &  &  &  \\
  &  &  & $^{182-184,186}$W($^{74-72,70}$Zn,$\alpha$)  &  &  &  &  \\
  &  &  & $^{182-184,186}$W($^{75-73,71}$Zn,$\alpha$n)  &  &  &  &  \\
\hline $^{272}$112 & $\sim$5x10$^{2}$$^{e}$ & 1.3x10$^{-15}$ &
$^{182-184,186}$W($^{90-88,86}$Sr,$\gamma$) & 2.0--9.8$^{f}$ &
1.8--0.05$^{g}$ & 5x10$^{-5}$ & 3.8 \\
\end{tabular}
\end{center}
\end{table}
\noindent$^{a}$ Excitation energies of the compound nuclei
produced by the $^{182}$W(X,p) (left) and the $^{186}$W(X,p)
(right) reactions are given, assuming bombarding energies which
are equal to the Coulomb barriers between the corresponding
projectile and target nuclei (r$_{0}$ = 1.4 fm).\\ $^{b}$ Fragment
cross sections were estimated according to the data of Ref.~[26]
(see text).\\ $^{c}$ The value of "f", the ratio of the production
cross section for a high energy fragment (E$_{Lab}$
$\stackrel{>}{\sim}$ 4.2 MeV/nucleon) to the total production
cross section of the particular fragment, was assumed to be 1\%
for the fragments which produced the $^{236}$Am isotope, and
reduced exponentially to 0.1\% for those which produced the
$^{252}$No isotope (see text).  \\ $^{d}$ The same cross section
was assumed for the various heavy ion reactions which contribute
to the production of a particular isotope.\\ $^{e}$ Ref.~[5].\\
$^{f}$ As "a" above but for the $^{182}$W($^{90}$Sr,$\gamma$)
(left) and $^{186}$W($^{86}$Sr,$\gamma$) (right) reactions. The
mass of the $^{272}$112 nucleus was taken from S.~Liran and
N.~Zeldes, {\it At. Data Nucl. Data Tables {\bf 17}, 431
(1976)}.\\ $^{g}$ Taken from U.~Trabitzsch and K.~B\"{a}chmann,
Radiochim. Acta {\bf 16}, 15 (1971).

\begin{figure}
\begin{center}
\leavevmode \epsfysize=12.0cm \epsfbox{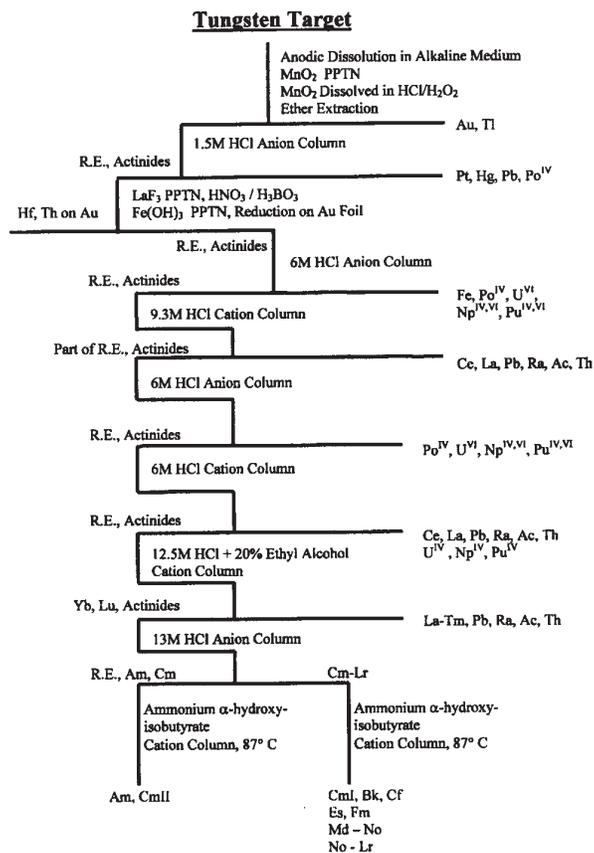}
\end{center}
\caption{Block diagram of the chemical separation of the actinide
fraction
 from the W target.}
\end{figure}
\newpage
\begin{figure}
\begin{center}
\leavevmode \epsfysize=8.5cm \epsfbox{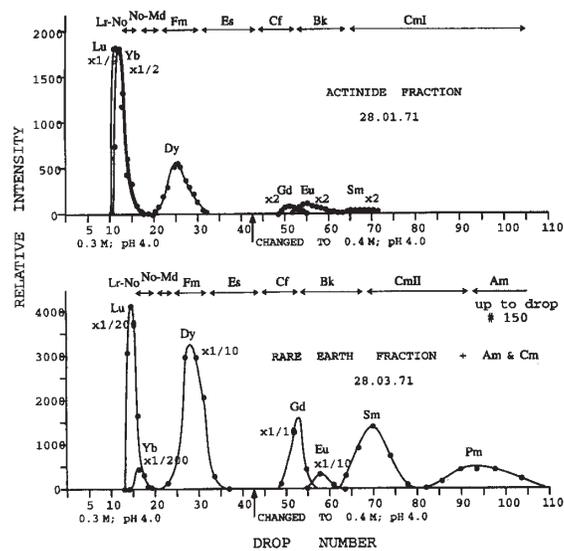}
\end{center}
\caption{Relative elution positions of rare earth elements from
Dowex-50 with pH 4.0 ammonium $\alpha$-hydroxyisobutyrate at
87$^{o}$ C. Top: Actinide fraction. Bottom: Rare earth fraction +
Am and Cm.}
\end{figure}
\begin{figure}
\begin{center}
\leavevmode \epsfysize=8.0cm \epsfbox{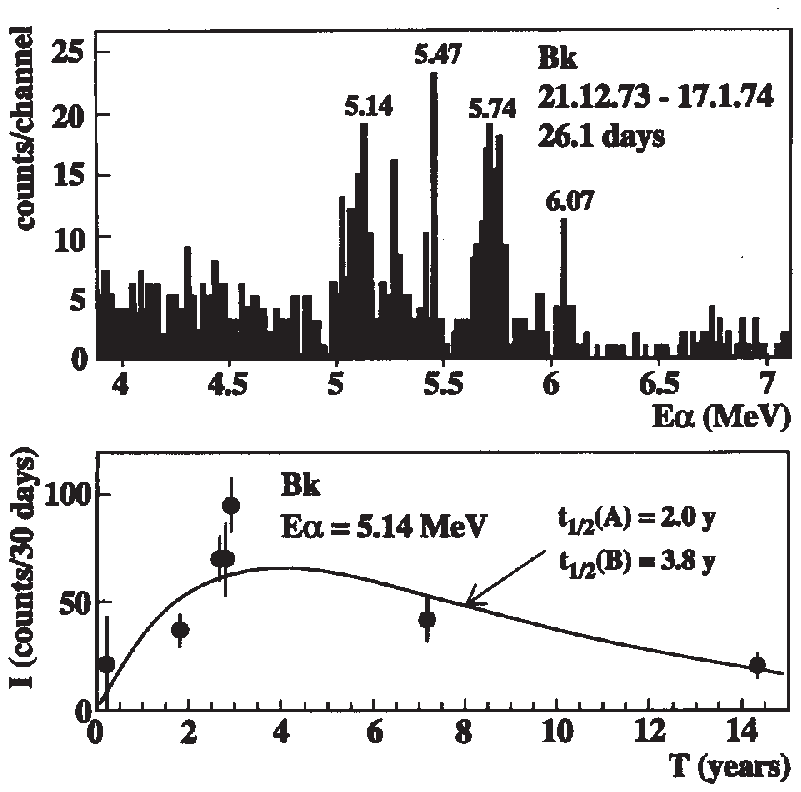}
\end{center}
\caption{Top: $\alpha$-particle spectrum obtained with the Bk
source. Bottom: Decay curve of the 5.14-MeV group shown in  the
top figure.}
\end{figure}
\begin{figure}
\begin{center}
\leavevmode \epsfysize=8.0cm \epsfbox{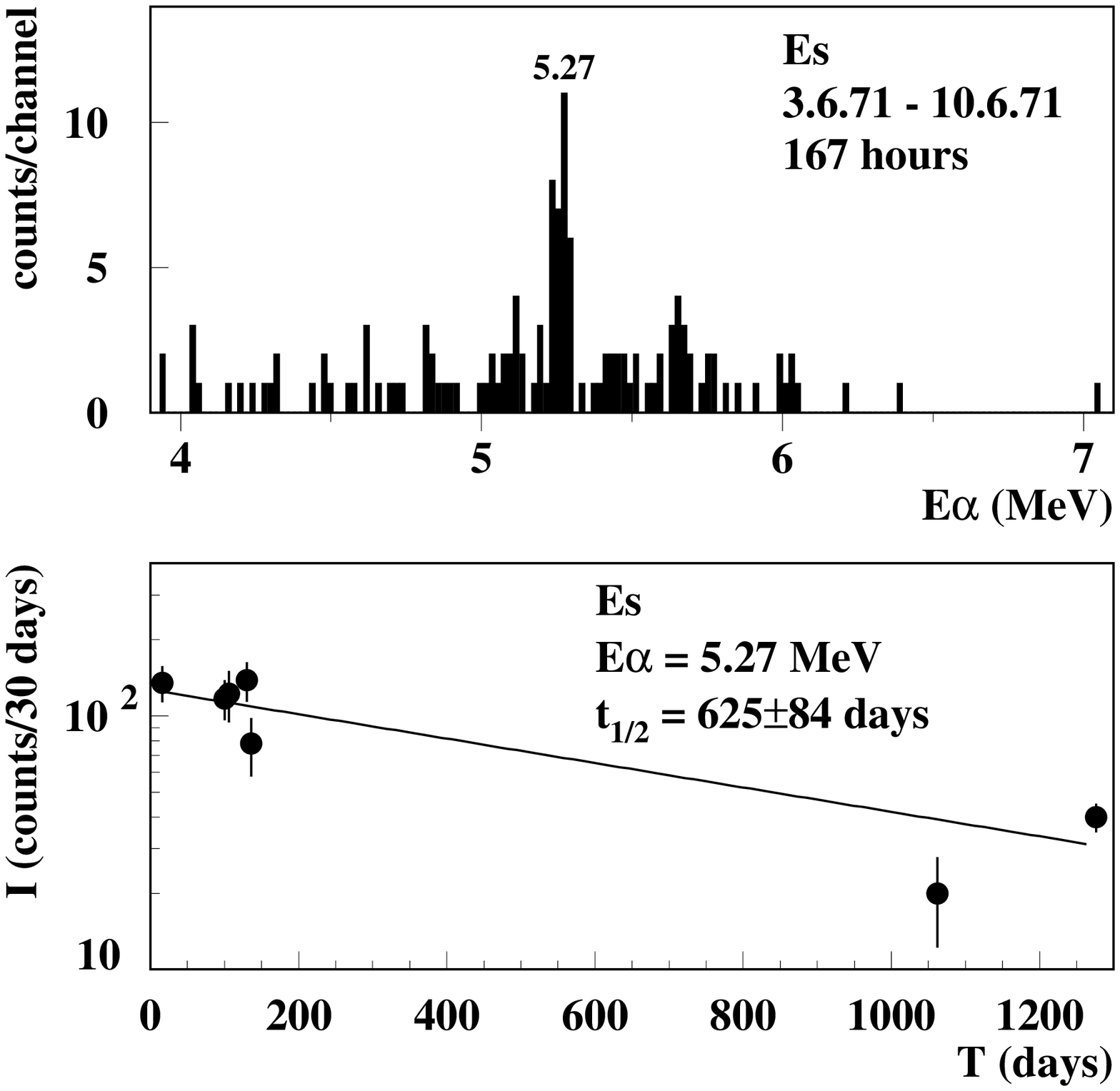}
\end{center}
\caption{Top: $\alpha$-particle spectrum obtained with the Es
source. Bottom: Decay curve of the 5.27-MeV group shown in the top
figure.}
\end{figure}
\begin{figure}
\begin{center}
\leavevmode \epsfysize=8.0cm \epsfbox{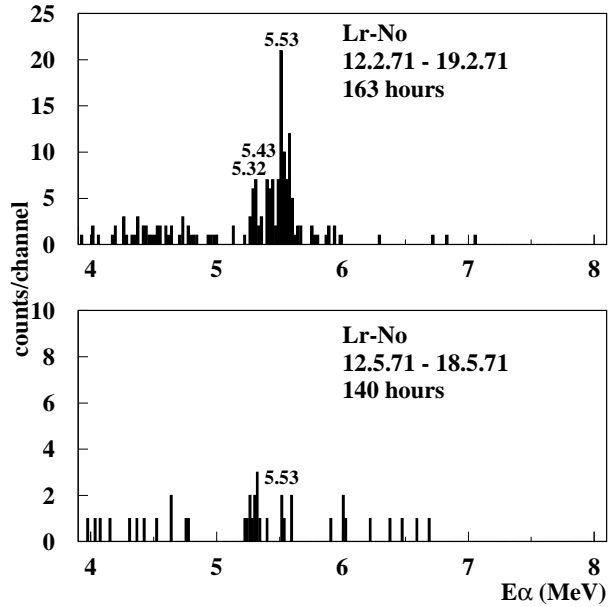}
\end{center}
\caption{Top: $\alpha$-particle spectrum obtained with the Lr-No
source. Bottom: The same as above but taken about 3 months later.}
\end{figure}
\begin{figure}
\begin{center}
\leavevmode \epsfysize=8.0cm \epsfbox{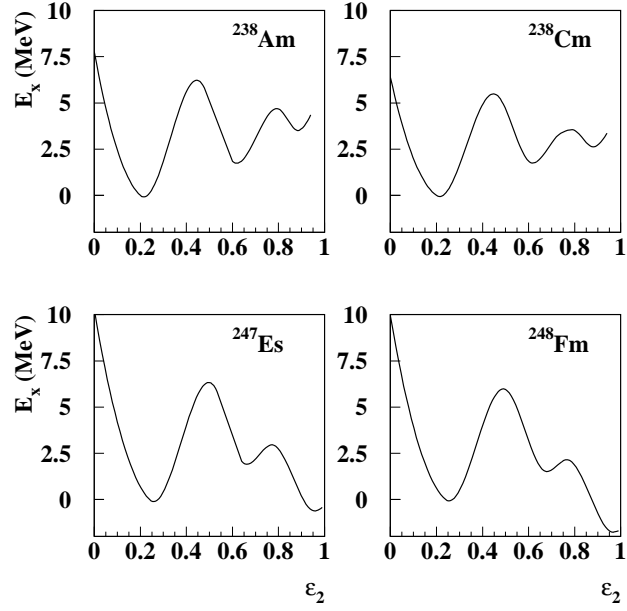}
\end{center}
\caption{Potential energies as function of quadrupole deformations
for 4 nuclei according to ref. [12].}
\end{figure}
\begin{figure}
\begin{center}
\leavevmode \epsfysize=8.0cm \epsfbox{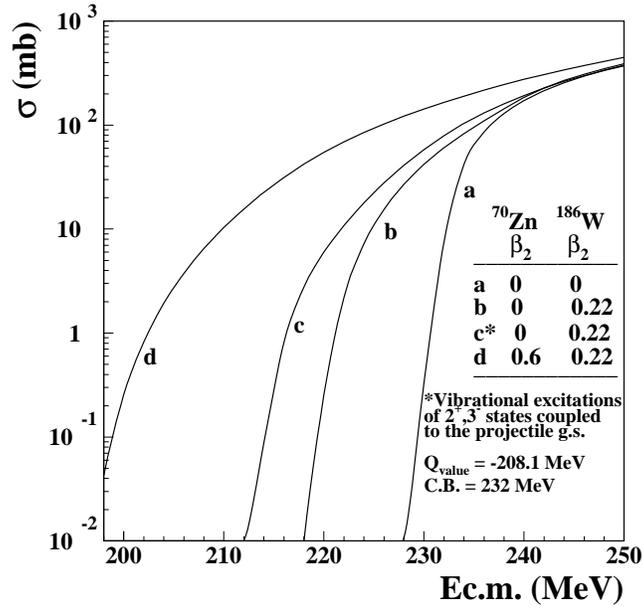}
\end{center}
\caption{Calculated fusion cross sections using the Code CCDEF
[28] for the $^{70}$Zn + $^{186}$W reaction assuming various
quadrupole deformations of the projectile and target nuclei (see
text).}
\end{figure}
\begin{figure}
\begin{center}
\leavevmode \epsfysize=5.0cm \epsfbox{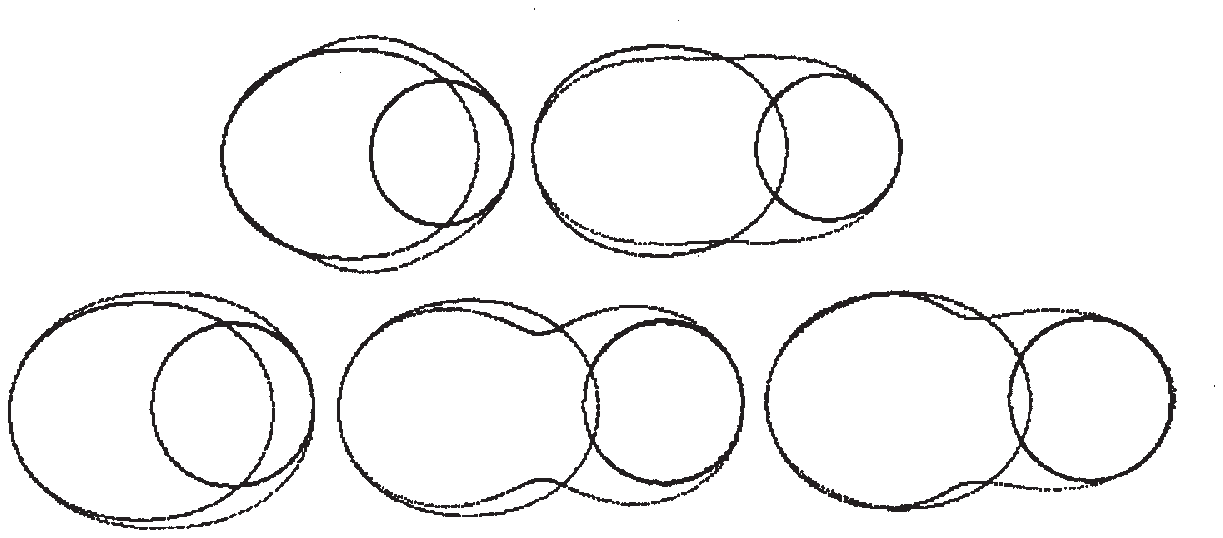}
\end{center}
\caption{Calculated shapes of two compound nuclei at various
configurations together with the shapes of the corresponding
projectile and target nuclei. Top, left: A$_{C.N.}$ = 239 in the
normal ground state; $\beta$$_{2}$ = 0.2; $\beta$$_{4}$ = 0.08
[12]. Top, right: A$_{C.N.}$ = 239 in the second minimum;
$\beta$$_{2}$ = 0.77; $\beta$$_{4}$ = 0.1 [12]. In both figures:
A$_{heavy}$ = 186; $\beta$$_{2}$ = 0.22 [29]. A$_{light}$ = 53;
$\beta$$_{2}$, $\beta$$_{3}$, $\beta$$_{4}$ = 0.0. Bottom, left:
A$_{C.N.}$ = 253 in the normal ground state; $\beta$$_{2}$ = 0.28;
$\beta$$_{4}$ = 0.01 [12]. Bottom, center: A$_{C.N.}$ = 253 in the
third minimum; $\beta$$_{2}$ = 1.2; $\beta$$_{4}$ = 0.0 [12].
Bottom, right: A$_{C.N.}$ = 253 with parameters of the third
minimum of $^{232}$Th; $\beta$$_{2}$ = 0.85; $\beta$$_{3}$ = 0.35;
$\beta$$_{4}$ = 0.18 [25]. In the three figures at the bottom:
A$_{heavy}$ = 186; $\beta$$_{2}$ = 0.22 [29]. A$_{light}$ = 67;
$\beta$$_{2}$, $\beta$$_{3}$, $\beta$$_{4}$ = 0.0.}
\end{figure}
\end{document}